\documentclass[11pt]{article}
\usepackage{amsmath,epsfig,epsf,psfrag}
\usepackage{graphicx,color}
\usepackage[top=0.5in, bottom=0.75in, left=0.5in, right=0.5in]{geometry}
\usepackage{setspace}

\title{A note on logistic regression and logistic kernel machine models}

\author{
Ru Wang$^{1}$, Jie Peng$^{1}$, Pei Wang$^{2}$\\
\\
$^1$ Department of Statistics, University of California, Davis, CA, 95616\\
$^2$ Division of Public Health Sciences, Fred Hutchinson Cancer
Research Center, Seattle, WA, 98109.
}
\date{}

\begin{document}

\maketitle

This is a note on logistic regression models and logistic kernel machine models. It contains derivations to some of the expressions in \cite{Wang}. \\

\noindent \textbf{\textit{Logistic regression models and score
tests}} \hspace{10pt} \\For the $i$th individual ($i=1,\dotsi, n$),
let response $y_i$ be 0 if unaffected, and 1 if affected. Let $X_i$ be a $q\times 1$ covariates vector (including an
intercept term), $z_i$ be a $p\times1$ vector of  SNP genotypes (or
summary scores) for a given gene (SNP set) under testing, and $s_i$
be the environment covariate which is also included in $X_i$.  We
consider the logistic regression model with gene-environment
interactions,
\begin{equation}\label{eq:logistic}
logit(p_i)=X^T_i\beta+\mathbf{a}^Tz_i+s_i\cdot \mathbf{b}^Tz_i,
~~~i=1,\cdots,n,
\end{equation}
where $p_i=Pr(y_i=1|X_i,z_i)$. The goal is to test the null
hypothesis $H_0: \mathbf{a}=\mathbf{b}=\mathbf{0}$. Consider  the score
statistic,
\begin{equation}\label{eq:score}
SS=\left((Y-\mu^0)^TZ^T, (S\cdot Y-S\cdot
\mu^0)^TZ^T\right)I^{22}\binom{Z(Y-\mu^0)}{Z(S\cdot Y-S\cdot \mu^0)},
\end{equation}
where $Y=(y_1,\cdots,y_n)^T,
S=(s_1,\cdots,s_n)^T,Z^T=(z_1^T,\cdots,z_n^T)$, and ``$\cdot$"
stands for the element-wise multiplication. In addition,
$$\mu^0=(p^0_1, \cdots, p^0_n)^T,$$ where $p^0_i$'s are
the fitted values of $p_i$'s under $H_0$. The information matrix
$$I^{22}=(I_{22}-I_{21}I^{-1}_{11}I_{12})^{-1},$$ where
\[
I_{11}=XD_1X^T, ~~ I_{12}=(ZD_1X^T, ZD_2X^T)=I_{21}^T, ~~I_{22} =
\begin{pmatrix}
ZD_1Z^T & ZD_2Z^T\\
ZD_2Z^T & ZD_2Z^T
\end{pmatrix},
\]
and $D_1$=diag$(p^0_i(1-p^0_i))$, $D_2$= diag$(s_i\cdot
p^0_i(1-p^0_i))$.
Under $H_0$, $SS \sim
\chi^2_{(n-v)}$, where $\nu$ is the rank of the matrix $(Z, S\cdot Z)$, where $S\cdot Z:=(s_1\cdot z_1^T,\cdots,s_n\cdot z_n^T)^T$. \\

\textbf{\textit{Logistic kernel machine models}} \hspace{10pt}\\
Following \cite{Liu,Wu}, we now extend (\ref{eq:logistic}) to a
semiparametric logistic regression model
\begin{equation}\label{eq:logistic.kernel}
logit(p_i)=X^T_i\beta+h(z_i)+s_i \cdot g(z_i), ~~~i=1,\cdots, n,
\end{equation}
where $h(\cdot)$ and $g(\cdot)$ belong to \textit{reproducing kernel
Hilbert spaces} $\mathbf{H}_K$ and $\mathbf{H}_{\widetilde{K}}$
generated by kernels $\mathbf{K}(\cdot, \cdot)$ and
$\widetilde{\mathbf{K}}(\cdot, \cdot)$, respectively. Considering penalized likelihood, $h(\cdot)$ and $g(\cdot)$ can be estimated by
\begin{equation}
(\hat{h}, \hat{g})=\textrm{argmax}_{h \in \mathbf{H}_K, g \in \mathbf{H}_{\widetilde{K}}}\left\{\sum_{i=1}^n (y_i
log(\frac{p_i}{1-p_i})+log(1-p_i))-\frac{1}{\lambda}\|h\|^2_{\mathbf{H}_K}-\frac{1}{\widetilde{\lambda}}\|g\|^2_{\mathbf{H}_{\widetilde{K}}}\right\}.
\end{equation}
Following \cite{Liu}, the above solutions have the same form as the
Penalized Quasi-Likelihood estimators from the logistic mixed model:
\begin{equation}\label{eq:logistic.random}
logit(p_i)=X^T_i\beta+h_i+s_i \cdot g_i, ~~~i=1,\cdots,n,
\end{equation}
where $h_i\sim_{i.i.d.} N_n(0, \frac{1}{\lambda}K), g_i\sim_{i.i.d.}
N_n(0,\frac{1}{\tilde{\lambda}}\tilde{K})$, and $h_i$'s and $g_i$'s are independent. Denote $\tau=1/\lambda$ and
$\tilde{\tau}=1/\widetilde{\lambda}$. Now, testing the
null hypothesis of no genetic effects $H_0: h(\cdot)=g(\cdot)=0$ in
(\ref{eq:logistic.kernel}) can be reformulated as testing the absence of the variance components $H_0: \tau=\tilde{\tau}=0$ in model (\ref{eq:logistic.random}). As in \cite{Liu, Wu}, we consider the
following test statistic based on the score statistic of $(\tau, \tilde{\tau})$:
\begin{equation}
Q=\binom{Q_{\tau}}{Q_{\tilde{\tau}}}=\binom{\frac{1}{2}(Y-\mu^0)^TK(Y-\mu^0)}{\frac{1}{2}(Y-\mu^0)^T(s\tilde{K})(Y-\mu^0)},
\end{equation}
where the $n \times n$ matrices $K:=(K(z_i,z_j)), \tilde{K}:=(\tilde{K}(z_i,z_j)), s\tilde{K}:=(s_i s_j\tilde{K}_{ij})$. Under $H_0$, the mean and
covariance of $Q$ are,
\[
\mu=\binom{\mu_{\tau}}{\mu_{\tilde{\tau}}}=\binom{\frac{1}{2}tr(P_0K)}{\frac{1}{2}tr(P_0s\tilde{K})},
~~~ I_{Q} =
\begin{pmatrix}
\sigma^2_{\tau} & \sigma_{\tau\tilde{\tau}}\\
\sigma_{\tilde{\tau}\tau} & \sigma^2_{\tilde{\tau}}
\end{pmatrix}
=
\begin{pmatrix}
\frac{1}{2}tr(P_0KP_0K) & \frac{1}{2}tr(P_0KP_0s\tilde{K})\\
\frac{1}{2}tr(P_0s\tilde{K}P_0K) &
\frac{1}{2}tr(P_0s\tilde{K}P_0s\tilde{K})
\end{pmatrix}
\]
where $$P_0:=W_0-W_0X(X^TW_0X)^{-1}X^TW_0,$$ and
$W_0=diag(p_i^0(1-p_i^0))$. We then linearly transform $Q$ to make its
two components uncorrelated:
 $$Q^*=\binom{Q^*_{\tau}}{Q^*_{\tilde{\tau}}}=I^{-\frac{1}{2}}_Q
Q$$ and $\mu^*=I^{-\frac{1}{2}}_Q \mu$. Since the components of
$Q^{\ast}$ are quadratic forms, they can be approximated by scaled
chiquare distributions  $\kappa^*_\tau \chi^2_{(\nu^\ast_\tau)},
\kappa^*_{\tilde{\tau}} \chi^2_{(\nu^\ast_{\tilde{\tau}})}$,
respectively \cite{Liu}. Through matching the means and variances, we
have $k^*_{\tau}=1/(2\mu^*_{\tau})$,
$k^*_{\tilde{\tau}}=1/(2\mu^*_{\tilde{\tau}})$ and
$\nu^*_{\tau}=2\mu^{*2}_{\tau}$,
$\nu^*_{\tilde{\tau}}=2\mu^{*2}_{\tilde{\tau}}$. Finally, we
construct a combined test statistic
\begin{equation}
Q^*_{max}=max(\frac{Q^*_{\tau}}{k^*_{\tau}},
\frac{Q^*_{\tilde{\tau}}}{k^*_{\tilde{\tau}}})
\end{equation}
The corresponding p-value is then
\[
\textrm{p-value}=1-F_{\chi^2}(Q^*_{max},\nu^*_{\tau})\cdot
F_{\chi^2}(Q^*_{max},\nu^*_{\tilde{\tau}}),
\]
where $F_{\chi^2}(\cdot, \nu)$ is the cumulative distribution function of a
chisquare distribution with $\nu$ degrees of freedom. Note, when both $K$
and $\tilde{K}$ are linear kernels, i.e., $K(z_i,z_j)=\widetilde{K}(z_i,z_j)=z_i^{T}z_j$,
models (\ref{eq:logistic}) and (\ref{eq:logistic.kernel}) have the same form. However, they are treated differently and consequently the corresponding test statistics
are different.
\\

\end{document}